\documentclass[english,aps,12pt,manuscript,showpacs,floats]{revtex4}
\usepackage{graphicx,epsfig}
\usepackage[latin1]{inputenc}
\usepackage{babel}
\usepackage[dvips]{color}

\makeatletter

\def\nn{\nonumber \\}

\def\p{\phi}
\def\pd{\dot\phi}
\def\be{\begin{equation}}
\def\ee{\end{equation}}
\newcommand{\brr}{\begin{eqnarray}}
\newcommand{\err}{\end{eqnarray}}

\makeatother
\begin{document}

\title{Cosmological stabilization of moduli with steep potentials}
\author{R. Brustein\protect\( ^{(1)}\protect \),
S. P. de Alwis\protect\( ^{(2)}\protect \), P. Martens\protect\( ^{(2)}\protect \)}
\affiliation{(1) Department of Physics, Ben-Gurion University, Beer-Sheva 84105,
Israel
\\
(2) Department of Physics,  University of Colorado, Box 390, Boulder, CO 80309.\\
 \texttt{e-mail:  ramyb@bgu.ac.il, dealwis@colorado.edu, martens@colorado.edu} }
\begin{abstract}
A scenario which overcomes the well-known cosmological overshoot problem associated
with stabilizing moduli with steep potentials in string theory is proposed. Our
proposal relies on the fact that moduli potentials are very steep and that
generically their kinetic energy quickly  becomes dominant. However, moduli kinetic
energy  red-shifts faster than other sources when the universe expands.  So, if
any additional sources are present, even in very small amounts, they will
inevitably become dominant. We show that in this case cosmic friction allows the
dissipation of the large amount of moduli kinetic energy that is required for the
field to be able to find an extremely shallow minimum. We present the idea using
analytic methods and verify  with some numerical examples.
\end{abstract}


\preprint{COLO-HEP 490}
\maketitle

\section{Introduction}

Let us start with a brief review of the framework in which the problem of
stabilizing string moduli in the perturbative outer region of moduli space, where the string
coupling is weak and/or the volume of the compact dimensions is large, is posed.

Realistic models, for example in flux compactifications of string theory,
usually include an effective $N=1$ supergravity (SUGRA) theory below the string
scale $M_s=10^{-2}$ (our conventions are such that $M_p\equiv\frac{1}{8\pi
G_N}=1.2\times 10^{18} GeV=1$) and supersymmetry (SUSY) breaking in some  hidden
sector at an intermediate scale $M_I=10^{-7}$. General arguments based on
symmetries show that the moduli superpotential must be a sum of exponentials in the
moduli and perhaps an additional constant. These exponentials could be generated by
stringy or field theoretic non-perturbative effects.

In this framework, a typical potential for moduli fields $\sigma_i$ has the $N=1$ SUGRA
form:
\begin{equation}
 \label{vsugra}
V_{SUGRA}=e^{K}\left(K^{i\overline{j}}F_{i}F_{\overline{j}}-3|W|^{2}\right),
\end{equation}
and a typical superpotential has the form:
 \be
 W(\sigma )=\sum_i A_ie^{-\alpha_i\sigma}.
 \ee
Where $K^{i\bar{j}}$ is the inverse metric derived from the Kahler potential $K$
and $F_i=\partial_i W+\partial_i K W$ is the Kahler derivative. Here we will
consider a situation where all but one of the moduli have been stabilized at the
string scale so that we can focus on the dynamics of one light modulus $\sigma$ as in the
recent work of Kachru et al (KKLT)\cite{Kachru:2003aw}.

The Kahler potential of moduli in the perturbative region is typically logarithmic
in the fields, so that in terms of a canonically normalized component field $\phi$
the potential involves exponentials of exponentials. An important feature of such
potentials is that they vanish or approach a finite constant when
$\phi\rightarrow\infty$.  This corresponds to decompactification if $\phi$ is the
volume modulus and zero coupling if $\phi$ is the effective dilaton. This is the
well known Dine-Seiberg problem \cite{Dine:1985he}. Therefore, if the potential has
a minimum, it needs to be separated from the asymptotic region by a potential
barrier.

Realistic models need some fine tuning of parameters. SUSY breaking in the
observable sector at the TeV scale requires that at the minimum of the potential
$F_{min}=M_I^2=O(10^{-14})$. In addition, if one assumes that the recent cosmological
observations indeed indicate that the cosmological constant (CC) is non-vanishing
then at the minimum $V_{min}=O(10^{-120}) > 0$\footnote{This assumption will not be
very important for us. All our considerations go through almost unmodified if
$V_{min}=0$.}, so $|F_{min}|=\sqrt{3}|W_{min}|+O(10^{-120})$. A stable minimum further
requires tuning of at least 4 parameters to prevent tachyonic directions. Until
recently it was very hard to find a single working model because the framework and
parameters were too constrained. Now, with the development of models based on flux
compactifications and the understanding of their vast parameter space, the
discretuum, it has become possible to find models with minima in the outer region
of moduli space \cite{Kachru:2003aw}.

\begin{figure}
\begin{center}
{\epsfig{file=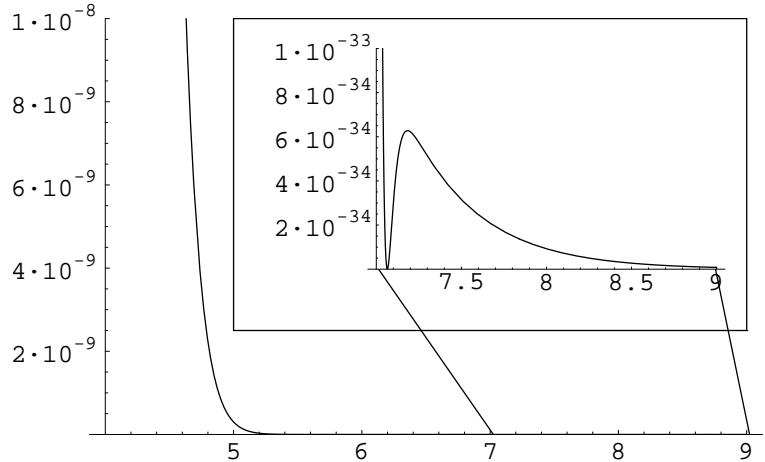,width=10.0cm}}
\end{center}
\caption{A typical moduli potential. The region of the shallow minimum had to be
magnified by 26 orders of magnitude so that it can be seen. The vertical axis is in
units of $M_p^4$, and the horizontal axis is in units of $M_p$.} \label{F1}
\end{figure}

We are interested in estimating the height of the barrier that separates the minimum
of the potential from the asymptotic region where the potential vanishes or
approaches a constant. Let us assume that at some value $\phi_{min}$ the potential has
a true minimum. Since each  exponential term is smaller in absolute value for
$\phi
> \phi_{min}$, we can generically expect that for $\phi
>\phi_{min}$, $|F|<|F_{min}|$, and $|W|<|W_{min}$ so that the height of the separating barrier
is at most limited by the intermediate scale $V_{max}\sim M_I^4=10^{-28}$. If no
further tuning is performed the height of the barrier is much lower than this
estimate. Typical moduli potentials are therefore steep and have a very shallow
minimum, when a stable one exists. A typical steep moduli potential with a shallow
minimum is shown in Fig.~\ref{F1}.

If one considers time-dependent solutions one encounters a cosmological version of
the Dine-Seiberg problem, first discussed by Brustein and Steinhardt \cite{bs}.  If
the moduli start from a generic point on the potential they are expected to reach
the outer region of moduli space by classically rolling towards the asymptotic
region. If they start to the left of the minimum, they will roll over the shallow
barrier, and if they start to the right of the barrier, they will never reach the
minimum and roll to the asymptotic outer region. To avoid this without additional
sources, such as radiation, the initial position of the field, $\phi_0$, has to be
such that the initial height of the field, $V(\phi_0)$, is at most an order of
magnitude larger than the height of the barrier. Thus getting a bound solution
requires fine-tuning of initial conditions to a very high accuracy. In addition,
the steep potential inhibits the possibility of inflation while the moduli are
rolling, since moduli kinetic energy tends to dominate the energy budget of the
universe. The consideration of time-dependent solutions thus leads to additional
criteria of stability of moduli beyond the standard static stability criteria, and
therefore leads to additional requirements and constraints beyond the standard
criteria which guarantee the absence of tachyonic directions. Such arguments have
led us to propose previously that the most likely place for moduli stabilization is
the central region of moduli space \cite{Brustein:2000mq}.

Several previous attempts to resolve the cosmological stability of moduli in
general, and in particular the cosmological overshoot problem, have been made.
Banks et al. have emphasized the role of the non-zero modes \cite{bbsms} and have
noticed that they red-shift slower than zero modes.  Dine  \cite{dine} noticed the
possible role of such slower redshift in stabilizing moduli. Barriero et al.
\cite{barreiroetal} discovered that the presence of additional sources helps to
relax the problem, however not at the level that is required.  Furthermore Huey et
al. \cite{hueyetal}  determined that if the moduli receive some specific
temperature corrections then the other sources (ie. radiation) could be much more
effective.  We will comment on this further below. Of course all these discussions
were made in the context of then known string models which were much more
constrained than the models that we have now.

Our proposed resolution relies  on the existence of other sources, for example, a
gas of relativistic particles: radiation. The nature of the additional sources and
their energy density is not particularly important to us.  A key fact that is
crucial to the scenario that we propose is that kinetic energy (KE) in an expanding
universe red-shifts at the fastest rate of all known sources. As the field rolls on
its steep potential its KE builds up and very quickly leads to KE dominance. Since
KE redshifts faster than the other sources, the additional sources will eventually
become the dominant energy component. While these sources dominate, they create a
large amount of cosmic friction which dissipates a large amount of energy, allows
the field to gently land into the basin of attraction of the shallow minimum, and
to eventually settle down at the minimum. While these additional sources dominate,
the field moves only a finite  amount \footnote{This fact was noticed by previous
authors in a different context \cite{skordis,linde}.},\footnote{Similar phases of
moduli evolution were discussed in a different context in \cite{kaloper}.}.

We do not rely on additional temperature dependent coupling beyond cosmic friction.
An example of such coupling is $T^2\phi^2$. In fact, in the particular example that
we use to illustrate our idea we assume that they are absent.  If such couplings
exist (as assumed by  Huey et al. \cite{hueyetal}) they could further help in
relaxing the constraints on moduli evolution. However, it has been argued that high
temperature effects do not modify the form of the potential of string theoretic
moduli \cite{binetruy}.

Since our main goal here is to explain our idea and show that it can be realized
rather than to explore in a general and systematic way the various possibilities
and caveats, we concentrate on some explicit examples that allow us to discuss the
basic argument. We will present a systematic search in a forthcoming publication.
In Section II we use approximate analytic solutions to study modular cosmology with
sources, and in Section III we verify numerically that our approximations and
quantitative estimates are valid. Section IV contains our conclusions and a brief
description of possible extensions of our scenario.

\section{Modular cosmology in the presence of sources}

We will discuss a specific model to expose the idea on which our proposal is based
and to explain its basic ingredients. We consider a cosmology with a single field
that has a potential of the form shown in Fig.~\ref{F1}. The field is assumed to
start in the steep region of the potential. Some amount of radiation in a thermal
state (with constant entropy so that $T\sim 1/a$) is also assumed to be present
$\rho_{rad}=C T^4=\frac{c^2}{a^4}$ where $C$ and $c$ are constants, $T$ is the
temperature and $a$ is the Friedman-Robertson-Walker (FRW) scale factor. We further
assume for simplicity a spatially flat universe. As we have explained, our proposal
is not particularly sensitive to the nature of the additional sources. In fact all
that is required is some additional source that redshifts slower than kinetic
energy (see below). Here we will just incorporate radiation - in a later paper we
will consider more general sources in detail. The qualitative features that we wish
to illustrate will however remain the same.

We do not discuss here the evolution prior to the epoch in which we can treat the
effective dynamics as a single scalar field in a cosmological background, or
whether the universe starts in a quantum region. In both of these cases the
universe would arrive at a starting point that is similar to the one that we
assume. Such initial conditions can be arrived at in different ways.
For example, a short period of inflation, a phase of so-called pre-big-bang
evolution, or nucleation from nothing.

The equations of motion that we need to solve are therefore
 \brr
H^2&=&\frac{1}{3}\left[\frac{\dot\phi^2}{2}+V(\phi ) +\frac{c^2}{a^4}\right],\nn
 \ddot\phi&+&3H\dot\phi+\frac{\partial V}{\partial\phi}=0.
 \err
Here a dot represents a derivative with respect to time, and $H$ is the Hubble
parameter $H=\dot a/a$.

Our scenario includes four  distinct epochs that are described below. Each epoch
starts with some specific initial values for the Hubble parameter and the field and
its derivatives. The end values of the previous epoch provide initial values for
the subsequent epoch. For definiteness, at the end of stage $i$ we denote the
values of the relevant variables by the subscript $i$ so for example $t_2$ is the
time at which epoch number 2 ends.

\begin{itemize}

\item

{\bf  epoch 1}: Potential domination

The field starts with zero initial velocity and a large potential energy on the
steep part of the potential. The universe expands at a fast rate. If a substantial
amount of radiation (or other sources) is initially present, the radiation energy
density quickly redshifts as $a^{-4}$, and the main source of energy becomes the
potential energy of the field.

The equations of motion can be approximated by
 \brr \ddot\phi +\frac{\partial V}{\partial\phi}&=&0 \nn
 H&=&\sqrt{\frac {V}{3}}.
\label{eqepoch1}
 \err
The energy $E=\frac{\dot\phi^2}{2}+V(\phi )$ is conserved in this epoch so all the
potential energy is converted into the field's kinetic energy (KE).

The solution of the approximate equations (\ref{eqepoch1}) with the initial
conditions at $t=0$, $\phi =\phi_0$, $V(\phi_0)=V_0$, and $\dot\phi=0$ is the
following,
 \brr
 \dot\phi&=&\sqrt{2\left[V_0-V(\phi)\right]} \nn
t&=&\int \frac{d\phi}{\sqrt{2\left[V_0-V(\phi)\right]}}\nn
 \ln a&=&\frac{1}{\sqrt 6}\int_{\phi_0}^{\phi}\frac{d\phi \sqrt{V(\phi )}}{\sqrt{V_0-V(\phi )}}.
\label{solepoch1}
 \err
The velocity of the field at the end of this epoch
 \be
\dot\phi_1=\sqrt{2\left[V_0-V(\phi_1)\right]},
 \label{dotphi1}
 \ee
can be quite large if the potential is steep $V_0\gg V(\phi_1)$, and the KE of the
field becomes the dominant energy component.

\item

 {\bf  epoch 2}: Kinetic energy domination

The equations of motion in this epoch can be approximated by
 \brr
 \ddot\phi+3H\dot\phi =0 \nn
 H^2=\frac{1}{6}\dot\phi^2.
\label{eqepoch2}
 \err

In epoch 2 the equation of motion for $\phi$ can be implicitly solved
$\dot\phi=\dot\phi_1a_1^3/a^3$, implying that $KE=\frac{1}{2}\dot\phi^2$ redshifts
quickly $KE=\frac{1}{2} \dot\phi_1^2 a_1^6/a^6$.

The solution of the approximate equations (\ref{eqepoch2}) with the initial
conditions at $t=t_1$, $\phi =\phi_1$, and
$\dot\phi=\dot\phi_1=\sqrt{2\left[V_0-V(\phi_1)\right]}$, given in
eq.(\ref{dotphi1}) is the following,
 \brr
\p-\p_1&=&\sqrt{\frac{2}{3}}\ln \left[\sqrt{\frac{3}{2}}\pd_1(t-t_1)+1\right] \nn
 a(t)&=&\left[\sqrt{\frac{3}{2}}a_1^3\pd_1(t-t_1)+a_1^3\right]^{1/3}.
 \label{solepoch2}
 \err

We can relate the amount of KE dissipation to the displacement in the field during
this epoch
 \be
 \ln\left(\frac{\dot\phi_1^2}{\dot\phi^2}\right)=\sqrt{6} (\phi-\phi_1),
 \ee
which means that to dissipate a few tens of orders of magnitude in KE, as required,
the field needs to move quite a bit in (reduced) Planck units to the extreme outer
region of moduli space. If the field moves a several Planck lengths, as is the case
in models that we have considered, then only a few orders of magnitude of KE are
dissipated, approximately one order of magnitude per planck length displacement of
the field.

As the universe expands, both the KE and the radiation energy density redshift.
Since the radiation redshifts at a slower pace, it will eventually ``catch up" with
the KE no matter how small its initial value. A possibility is that the potential
energy will become dominant before the radiation. Whether this happens or not depends on the
details of the model. In the successful cases (in the case of steep potentials) the
radiation becomes dominant first.

\item

{\bf  epoch 3}: Radiation domination

Epoch 2 leads to a radiation dominated epoch when the radiation ``catches up" with
the KE, $\left(\rho_{\text rad}\right)_2\gg \left(KE\right)_2$,
 \be
 \frac{c^2}{a_2^4}\gg\frac{\dot\phi_1^2}{2}\frac{a_1^6}{a_2^6}
 \ee

In epoch 3 the KE continues to redshift as
$K=\frac{\dot\phi_1^2}{2}\frac{a_1^6}{a^6}$, and the expansion of the universe is
faster, hence in this epoch KE is dissipated in a more efficient way.

The equations of motion in epoch 3 can be approximated by
 \brr
 \ddot\phi+3H\dot\phi =0 \nn
 H^2=\frac{1}{3}\frac{c^2}{a^4}.
\label{eqepoch3}
 \err

The solution of eqs.(\ref{eqepoch3}) is given by
 \brr
 a&=&\sqrt{\frac{2c}{\sqrt{3}}(t-t_2)+a_2^2)}\nn
 \p &=&\p_2+\frac{\sqrt{3}\pd_1 a_1^3}{ca_2}
 -\frac{\sqrt{3}\pd_1 a_1^3}{c}\frac{1}{\sqrt{\frac{2c}{\sqrt{3}}(t-t_2)+a_2^2}}.
\label{solepoch3}
 \err
The displacement of the field during this epoch can be expressed after some algebra
as follows:
 \brr
 \phi-\phi_2=\sqrt{6}
 \sqrt{\frac{\left(KE\right)_2}{\left(\rho_{rad}\right)_2}}
 \left(1-\frac{a_2}{a}\right)\le \sqrt{6}.
 \label{dsiplacementepoch3}
 \err

Therefore, even if this epoch continues indefinitely the field will only move a
finite distance \cite{skordis,linde}. KE on the other hand, continues to red-shift
as $1/a^6$, so contrary to epoch 2, the field can dissipate a lot of energy while
staying almost constant.

\item

{\bf  epoch 4}: Potential domination

Since the radiation energy density redshifts quickly, eventually, the potential
energy density no matter how small, will come to dominate  when
$V(\phi)\gg\frac{c^2}{a_2^4}$. In this epoch the equations of motion can again be
approximated by
 \brr \ddot\phi +\frac{\partial V}{\partial\phi}&=&0 \nn
 H&=&\sqrt{\frac {V}{3}}.
\label{eqepoch4}
 \err
At this point the solution of this system depends on the value of $\phi$, and
whether it ended up in the basin of attraction of a minimum, beyond the last
minimum, or still at a high point. If the transition from KE dominance to RD occurs
at a point that is close enough to the minimum (and to the left of the maximum),
then the field will be trapped even in a shallow minimum since it has lost a huge
amount of KE.

\end{itemize}

The conclusion of this analysis is that as the field acquires a large amount of KE
the latter seeds the elements of its own destruction by the lurking radiation. In
the KE domination epoch the field does not move much per order of magnitude of
dissipated KE, and in the radiation dominated epoch its motion is clearly bounded.
Thus if the potential is steep enough, so that enough kinetic energy is acquired in
the initial stage, and friction becomes important early on, and the difference in
elevation between the starting point and the minimum is large enough, then with
generic initial conditions the field would be bound.

Obviously the detailed realization of this scenario depends on the values of the
parameters in the potential and the nature of the specific sources.

\section{Numerical examples}

To verify that our approximations are meaningful, and to check their range of
validity, we have made a series of numerical investigations. We report here only
about some of them to illustrate some features of the scenario.
 \begin{figure}[b]
\begin{center}
{\epsfig{file=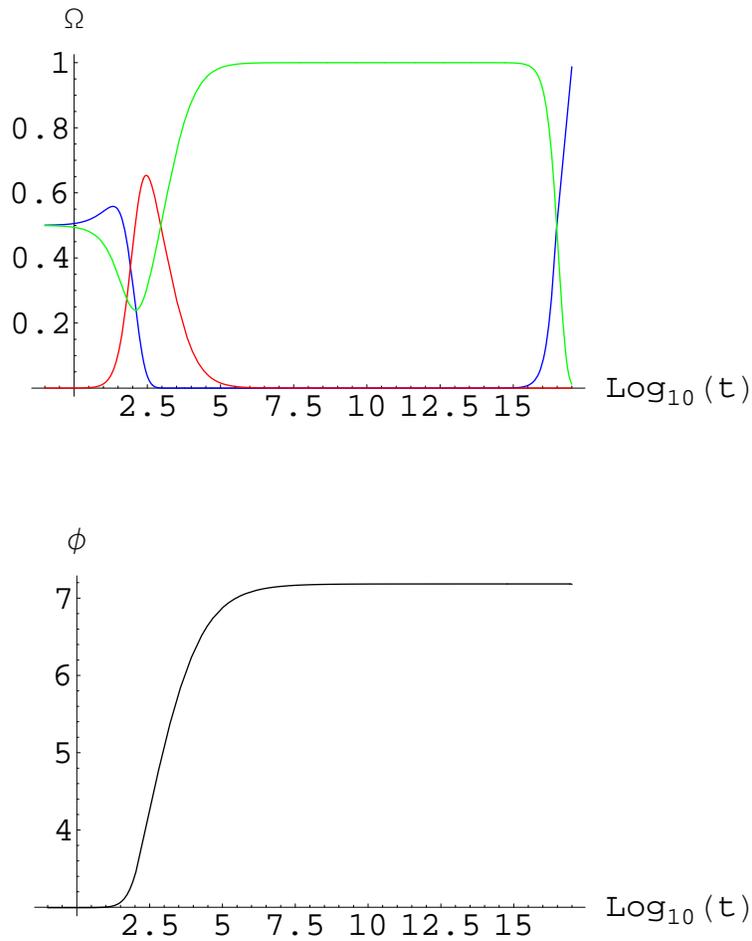,width=10.0cm}}
\end{center}
\caption{ A bound case. Shown in the top panel are fractional densities of
potential energy (blue), kinetic energy (red) and radiation energy (green) as a
function of time. KE becomes dominant and then the radiation. Shown in the bottom
panel is the evolution of the field as a function of time ending in the shallow
minimum of the potential. } \label{F2}
 \end{figure}

For the numerical investigation we have used the following potential for the volume
modulus used in \cite{Kachru:2003aw}:
\begin{equation}
\label{kkltpot}
 V\left(\sigma\right) =
\frac{a\,A\,e^{-a\,\sigma}}{2\,\sigma^{2}}\left(\frac{1}{2}\,\sigma\,a\,A\,e^{-a\,\sigma}
+ W_{0} +A\,e^{-a\,\sigma}\right)+\frac{d}{\sigma^{3}},
\end{equation}
where $\sigma \equiv e^{\sqrt{2/3}\phi}$. This potential was derived using the
superpotential (coming from flux contributions and non-perturbative effects)
\begin{equation}
W\left(\sigma\right) = W_{0} + A\,e^{-a\,\sigma},
\end{equation}
and the classical tree level Kahler potential. The last term in eq.(\ref{kkltpot})
comes from the contribution of an anti-Dbrane (Dbar brane). We stress that we use
this particular potential merely for the sake of illustration of our mechanism.
Similar results follow for potentials  which incorporate alternatives to the Dbar
brane term.

\begin{figure}[h]
\begin{center}
{\epsfig{file=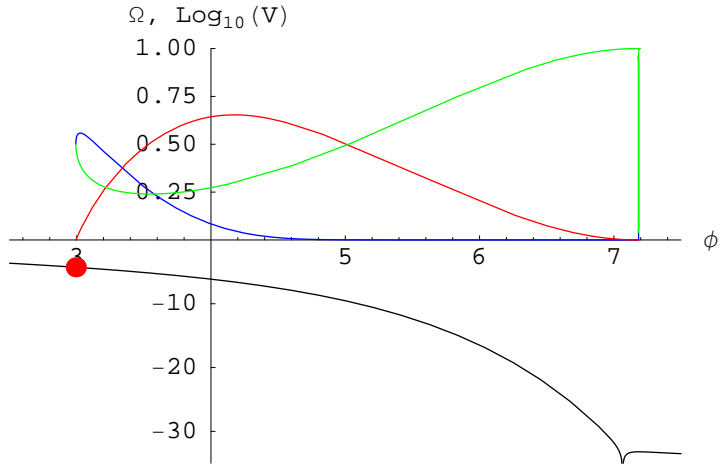,width=10.0cm}}
\end{center}
\caption{ A bound case. Shown in the top panel are the various energy densities
(color coded as in Fig.~\ref{F2}) as a function of the scalar field position. Shown
in the bottom panel is the potential and the starting position of the field. Note
that the scale is logarithmic and that the difference in potential energy between
the starting point and the minimum is about 30 orders of magnitude!  \label{F3} }
\end{figure}

\begin{figure}[t]
\begin{center}
{\epsfig{file=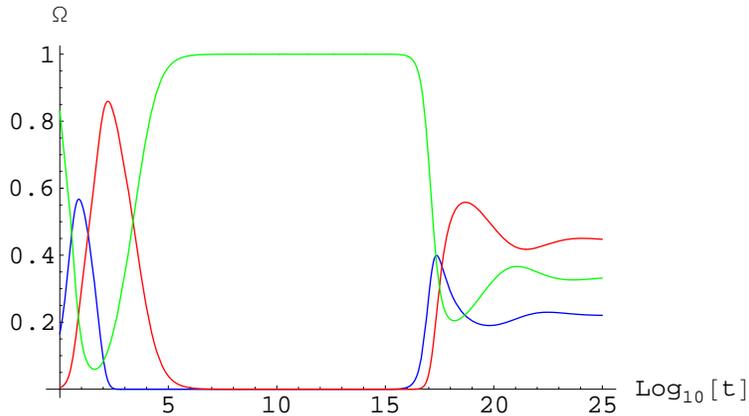,width=10.0cm}}
\end{center}
\caption{ An unbound case: various energy densities
(color coded as in Fig.~\ref{F2}) as a function of time. The only
difference from the bound case is where the field ends up at the
end of the radiation dominated epoch. In this case it lands to the
right of the shallow minimum and continues to run on the potential
in a ``tracking" solution.} \label{F4}
\end{figure}

\begin{figure}[t]
\begin{center}
{\epsfig{file=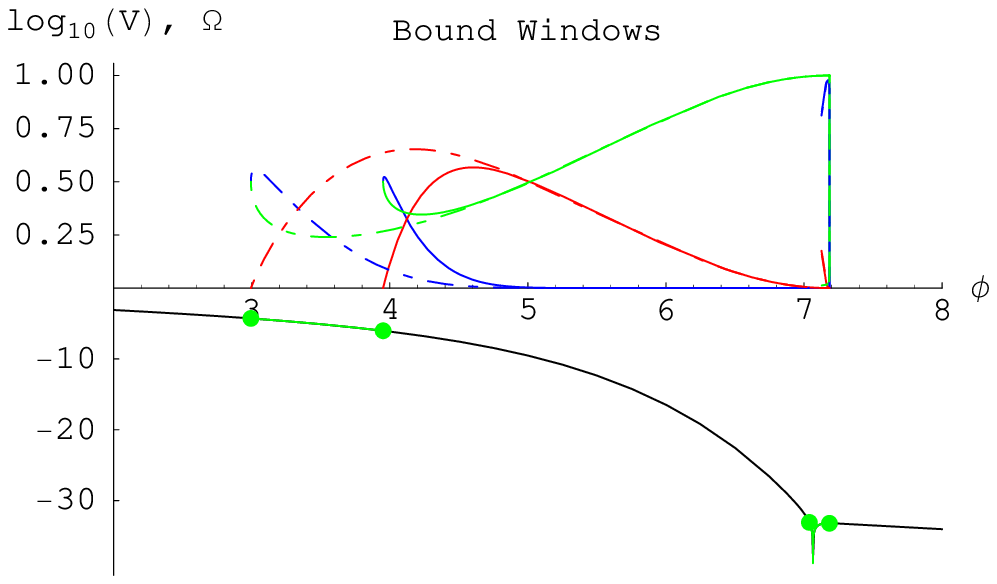,width=10.0cm}}
\end{center}
\caption{ Window of allowed initial conditions for bound
solutions. Shown in the top panel are the various energy densities
as a function of scalar field position (color coded as in
Fig.~\ref{F2}) for two bound solutions with two different initial
conditions, depicted by solid and dashed curves. In the lower
panel we show the potential as a function of the scalar field. The
green dots on the potential are the end points of the two regions
of initial conditions that lead to bound solutions. One of the
regions is around the minimum, and the other is way up on the
potential.} \label{F5}
\end{figure}

All our numerical investigations were done in the conventions that $M_{p}^{2}\equiv
\frac{1}{8\pi G_{N}} = 1$. We present our results in Figures \ref{F2} through
\ref{F5}.
To construct the numerical examples in these figures we used the
following values for our parameters:
 $a = 0.1$, $ A = 1.0$, $d = 3 \times 10^{-26}$ , $W_0 = -2.96\times 10^{-13}$.
For these parameters the potential has a  a true minimum at $\phi = 7.06$. At the
minimum the value of the potential (the CC) is $6.35\times 10^{-42}$ and the
barrier separating the minimum from the asymptotic region is located at $\phi =
7.18$. The height of the barrier is $6.28\times 10^{-34}$.

To illustrate the effect of radiation on the evolution of a moduli
field we present examples of a bound and an unbound case. These
two examples differ only in initial conditions.  To create these
two examples we set the initial conditions as follows. For the
bound case: $V_{0} = ({\rho_{rad}})_{0} = 5.20\times 10^{-5}
M_{p}^4$. The corresponding initial value of  the field is
$\phi_{0} = 2.99$, and the velocity of the field vanishes
initially. For the unbound case: $V_{0} = ({\rho_{rad}})_{0} =
3.78\times 10^{-2} M_p^4$. The fractional energy densities created
by our bound solution are shown in Figures \ref{F2} and \ref{F3}.
The first depicts the value of the field and the fractional energy
densities as a function of time, while the second shows the same
fractional energy densities and the potential as a function of the
field.  The four epochs mentioned previously in this paper are
clearly visible in both representations. The fractional energy
densities created by our unbound solution are shown in
Fig~\ref{F4}.  Note that a significant fourth epoch is not present
in this example. The field lands to the right side of the barrier,
where a scaling solution quickly takes over.

The addition of radiation to our model created an extra window,
$\left(2.99, 3.95\right)$, of initial conditions which lead to
bound solutions. The other interval, $\left(7.04, 7.18\right)$,
remains relatively unchanged with the addition of radiation.  The
two intervals of bound solutions can be seen in Fig~\ref{F5}.

\section{Discussion and outlook}

We have proposed a scenario for resolving the cosmological moduli stabilization
problem. We believe that we have identified the basic ingredients of the required
solution. The application of the idea may vary according to the detailed models of
string compactifications.

Obviously this preliminary investigation needs to be followed by detailed and
quantitative exploration, including various sources, a systematic study with a
wider range of parameters and potentials, and a quantitative analysis of stable
potentials.  In particular, the dependence of the range of bound initial conditions
on the parameters of the potential such as the width and height of the minimum and
barrier.

Having found cosmologically stable models, it is clear that string models become
less constrained. This provides a new perspective on the approach to string model
building. Previously, because the theory was highly constrained, the hope was that
only a single model, or a single class of models will satisfy the necessary
conditions. Now the theory needs to be constrained as much as possible by data and
phenomenological bottom-up constraints.

As a final comment we wish to point out that constructing models that lead to slow
roll inflation in this framework is rather easy to achieve. In models with
stabilized moduli, one needs to find some that have a flat enough barrier as first
discussed in the models of natural inflation \cite{natural}. Such a model in which
the inflaton is the axion associated with the volume modulus $\phi$ was recently
discussed in \cite{Blanco-Pillado:2004ns}. More  models of inflation from a
flat maximum of the potential of stabilized moduli were introduced in \cite{bda2},
and \cite{trivedi}. Additional regions in the discretuum may allow the construction
of other types of inflationary models.

\section{Acknowledgments}

This research is supported in part by the United States Department of Energy under
grant DE-FG02-91-ER-40672. We would like to thank C. Burgess, M. Dine, N. Kaloper
and A. Linde for useful comments and discussions.

\bibliographystyle{apsrev}

\begin{thebibliography}{10}
\bibitem{Kachru:2003aw}
S.~Kachru, R.~Kallosh, A.~Linde and S.~P.~Trivedi,
``De Sitter vacua in string theory,''
Phys.\ Rev.\ D {\bf 68}, 046005 (2003)
[arXiv:hep-th/0301240].

\bibitem{Dine:1985he}
M.~Dine and N.~Seiberg,
``Is The Superstring Weakly Coupled?,''
Phys.\ Lett.\ B {\bf 162}, 299 (1985).


\bibitem{bs}
R.~Brustein and P.~J.~Steinhardt,
``Challenges for superstring cosmology,''
Phys.\ Lett.\ B {\bf 302}, 196 (1993)
[arXiv:hep-th/9212049].



\bibitem{Brustein:2000mq}
R.~Brustein and S.~P.~de Alwis,
``String universality,''
Phys.\ Rev.\ D {\bf 64}, 046004 (2001)
[arXiv:hep-th/0002087].

\bibitem{bbsms}
T.~Banks, M.~Berkooz, S.~H.~Shenker, G.~W.~Moore and P.~J.~Steinhardt,
``Modular cosmology,''
Phys.\ Rev.\ D {\bf 52}, 3548 (1995)
[arXiv:hep-th/9503114].


\bibitem{dine}
M.~Dine,
``Towards a solution of the moduli problems of string cosmology,''
Phys.\ Lett.\ B {\bf 482}, 213 (2000)
[arXiv:hep-th/0002047].

\bibitem{barreiroetal}
T.~Barreiro, B.~de Carlos and E.~J.~Copeland,
``Stabilizing the dilaton in superstring cosmology,''
Phys.\ Rev.\ D {\bf 58}, 083513 (1998)
[arXiv:hep-th/9805005].

\bibitem{hueyetal}
G.~Huey, P.~J.~Steinhardt, B.~A.~Ovrut and D.~Waldram,
``A cosmological mechanism for stabilizing moduli,''
Phys.\ Lett.\ B {\bf 476}, 379 (2000)
[arXiv:hep-th/0001112].



\bibitem{skordis}
A.~Albrecht, C.~P.~Burgess, F.~Ravndal and C.~Skordis, ``Natural
quintessence and large extra dimensions,'' Phys.\ Rev.\ D {\bf
65}, 123507 (2002) [arXiv:astro-ph/0107573].

\bibitem{linde}
G.~N.~Felder, A.~V.~Frolov, L.~Kofman and A.~V.~Linde, ``Cosmology with negative
potentials,'' Phys.\ Rev.\ D {\bf 66}, 023507 (2002)
[arXiv:hep-th/0202017].

\bibitem{kaloper}
N.~Kaloper and K.~A.~Olive, ``Dilatons in string cosmology,''
Astropart.\ Phys.\  {\bf 1}, 185 (1993).


\bibitem{binetruy}
P.~Binetruy and M.~K.~Gaillard, ``Candidates For The Inflaton
Field In Superstring Models,'' Phys.\ Rev.\ D {\bf 34}, 3069
(1986).


\bibitem{natural}
F.~C.~Adams, J.~R.~Bond, K.~Freese, J.~A.~Frieman and
A.~V.~Olinto, ``Natural inflation: Particle physics models, power
law spectra for large scale structure, and constraints from
COBE,'' Phys.\ Rev.\ D {\bf 47}, 426 (1993)
[arXiv:hep-ph/9207245].


\bibitem{Blanco-Pillado:2004ns}
J.~J.~Blanco-Pillado, C.P. Burgess, J.M. Cline, C. Escoda, M. Gomez-Reino, R. Kallosh, A. Linde, F. Quevedo,
``Racetrack inflation,''
arXiv:hep-th/0406230.


\bibitem{bda2}
R.~Brustein, S.~P.~De Alwis and E.~G.~Novak, ``Inflationary
cosmology in the central region of string / M-theory  moduli
Phys.\ Rev.\ D {\bf 68}, 023517 (2003) [arXiv:hep-th/0205042].


\bibitem{trivedi}
N.~Iizuka and S.~P.~Trivedi, ``An inflationary model in string
theory,'' arXiv:hep-th/0403203.




\end{thebibliography}

\end{document}